\documentclass[11pt]{article}
\pdfoutput=1
\usepackage{soul}
\usepackage{jheppub}
\usepackage{amsfonts}
\usepackage{amsmath}
\usepackage{amsthm}
\allowdisplaybreaks[4]         
\usepackage{amssymb}   
\usepackage{euscript}   
\usepackage{cleveref}    
\usepackage[dvipsnames]{xcolor}          
\usepackage{tensor}     
\usepackage{graphicx}
\usepackage{caption}
\usepackage{subcaption}   
\usepackage{hyperref}
\hypersetup{
	colorlinks =NavyBlue,
	linkcolor =NavyBlue,
	citecolor=MidnightBlue,
	filecolor=NavyBlue,
	urlcolor=NavyBlue,
}
\usepackage{wasysym}
\usepackage{stmaryrd}

\newcommand{\req}[1]{(\ref{#1})} 

\newcommand{\bea}{\begin{eqnarray}}
\newcommand{\eea}{\end{eqnarray}}
\newcommand{\ba}{\begin{eqnarray}}
\usepackage{braket}
\newcommand{\ea}{\end{eqnarray}}

\newcommand{\beq}{\begin{equation}}
\newcommand{\eeq}{\end{equation} }
\newcommand{\beqa}{\begin{eqnarray}}

\newcommand{\eeqa}{\end{eqnarray}}
\newcommand{\beqar}{\begin{eqnarray*}}
\newcommand{\eeqar}{\end{eqnarray*}}

\newcommand{\be}{\begin{equation}}
\newcommand{\ee}{\end{equation}}
\newcommand{\diff}{\mathrm{d}}

\renewcommand{\req}[1]{(\ref{#1})}

\newcommand{\ie}{{\it i.e.,}\ }

\newcommand{\GN}{G_\mathrm{N}}

\definecolor{shadecolor}{rgb}{.25,.25,.25}

\title{ \boldmath Regular black holes from Oppenheimer-Snyder collapse }

\author[\lightning]{\small Pablo Bueno,}
\author[\lightning]{\small Pablo A.~Cano,}
\author[\scriptsize \sun]{\small Robie A.~Hennigar,}
\author[\lightning]{\small  \'Angel J.~Murcia,} 
\author[\lightning]{\small Aitor Vicente-Cano}

\affiliation[\lightning]{Departament de Física Quàntica i Astrofísica, Institut de Ciències del Cosmos\\ Universitat de Barcelona, Martí i Franquès 1, E-08028 Barcelona, Spain \vspace{0.1cm}}
\affiliation[\scriptsize \sun]{Centre for Particle Theory, Department of Mathematical Sciences\\ Durham University, Durham DH1 3LE, UK}

\emailAdd{pablobueno@ub.edu}
\emailAdd{pablo.cano@icc.ub.edu}
\emailAdd{robie.a.hennigar@durham.ac.uk}
\emailAdd{angelmurcia@icc.ub.edu}
\emailAdd{avicentecano@icc.ub.edu }

\date{\today}
\abstract{It has been recently shown that regular black holes arise as the unique spherically symmetric solutions of broad families of generalizations of Einstein gravity involving infinite towers of higher-curvature corrections in $D\geq 5$ spacetime dimensions. In this paper we argue that such regular black holes arise as the byproduct of the gravitational collapse of pressureless dust stars. We show that, just like for Einstein gravity, the modified junction conditions for these models impose that the dust particles on the star surface follow geodesic trajectories on the corresponding black hole background. Generically, in these models the star collapses until it reaches a minimum size (and a maximum density) inside the inner horizon of the black hole it creates. Then, it bounces back and reappears through a white hole in a different universe, where it eventually reaches its original size and restarts the process. Along the way, we study FLRW cosmologies in the same theories that regularize black hole singularities. We find that the cosmological evolution is completely smooth, with the  big bang and big crunch singularities predicted by Einstein gravity replaced by cosmological bounces. 
}

\begin{document} 
\vspace*{1\fill}
\maketitle
\flushbottom

\section{Introduction}

Gravitational collapse occurs ubiquitously in nature, probing all physical scales from the well-understood to the unknown. Stars resist collapse through the outward pressure generated by nuclear fusion reactions in their cores. Once their nuclear fuel is exhausted, stars with masses up to a few solar masses can persist due to quantum mechanical degeneracy pressure. For sufficiently massive stars, however, no known physical laws can halt their complete collapse. 

The first exploration of total gravitational collapse in the framework of general relativity was carried out by Oppenheimer and Snyder~\cite{Oppenheimer:1939ue}. Considering a spherically symmetric star composed of pressureless dust, they argued that the collapse causes the star to reach its Schwarzschild radius in a finite time. Though it was not clear to them at the time, their analysis in fact shows that the collapse proceeds beyond that point,  ultimately resulting in a spacetime singularity cloaked by an event horizon.
Their result was deeply inspirational in developing the modern understanding of gravitational collapse and black holes, which was put into its essential form by Penrose. Under quite general assumptions, Penrose proved that singularities are a \textit{generic} outcome of gravitational collapse~\cite{Penrose:1964wq}. He further conjectured that these singularities are always hidden inside the event horizon of a black hole, in a form of cosmic censorship~\cite{Penrose:1969pc}. Thus, though highly idealized, the Oppenheimer-Snyder model is expected to capture the qualitative features of generic gravitational collapse. 
%

The emergence of singularities in gravitational collapse presents a foundational problem. Our current physical theories break down at singularities, necessitating a more complete physical theory to describe physics in their vicinity. One source of hope lies in quantum properties of the gravitational field, which may evade some of the assumptions in Penrose’s theorem and become significant during the final stages of collapse. It is therefore widely expected that quantum gravity will tame and explain the singularities of general relativity. Unfortunately, no existing theory of quantum gravity is sufficiently developed to address the final fate of gravitational collapse.

In the absence of a complete top-down theory, a promising avenue is to extract general lessons from existing quantum gravity frameworks and implement them as robustly as possible within a bottom-up approach. Quantum gravity is expected to modify general relativity in several ways. On the one hand, genuine quantum fluctuations of the metric are anticipated to become significant near the Planck scale, eventually leading to a breakdown of the classical description of spacetime. These effects may ultimately resolve spacetime singularities, but accurately modeling them remains well beyond the reach of any current bottom-up construction. On the other hand, many approaches to quantum gravity predict that the Einstein–Hilbert action receives corrections at the classical level in the form of higher-derivative terms~\cite{tHooft:1974toh,Goroff:1985th,Zwiebach:1985uq,Gross:1986mw,Grisaru:1986vi,Sakharov:1967nyk,Visser:2002ew,Endlich:2017tqa,Eichhorn:2020mte,Borissova:2022clg}. The prototypical example is string theory, which predicts an infinite tower of higher-curvature terms; however, only the first few corrections are known explicitly. Capturing the effects of these terms in bottom-up frameworks is a difficult problem, but one that has recently been~addressed. 

Recently, three of us (BCH) have obtained exact, spherically symmetric black hole solutions to purely gravitational theories that incorporate infinite towers of higher-curvature corrections~\cite{Bueno:2024dgm}. This was achieved by working with \textit{quasi-topological gravities}~\cite{Oliva:2010eb,Quasi,Dehghani:2011vu,Ahmed:2017jod,Cisterna:2017umf,Bueno:2019ycr, Bueno:2022res, Moreno:2023rfl}, which are a class of higher-order gravitational theories that possess particularly desirable properties on spherically symmetric backgrounds. Specifically, quasi-topological gravities have second-order equations of motion in spherical symmetry and have a Birkhoff theorem, ensuring uniqueness of the static solution. While the actions of quasi-topological theories are carefully chosen to ensure these properties, the resulting theories are general enough to describe vacuum gravitational effective field theory in five and higher dimensions~\cite{Bueno:2019ltp}. Thus, there is good reason to believe the results obtained in this framework will capture some of the qualitative features of resumming infinite towers of higher-curvature corrections in top-down models.

%

The most striking feature of the solutions constructed in~\cite{Bueno:2024dgm} is that they are \textit{singularity-free}. Under very general conditions, the Schwarzschild singularity is replaced by a regular (anti)-de Sitter core in any dimension $D \ge 5$. This suggests that the classical corrections to general relativity predicted by various quantum gravity frameworks may be sufficient to resolve the singularities of Einstein's theory.\footnote{Some further existing hints of this possibility include regular black holes in higher-derivative theories with gauge fields~\cite{Cano:2020qhy,Cano:2020ezi,Bueno:2021krl,Bueno:2025jgc} and cosmological scenarios~\cite{Arciniega:2018tnn,Wang:2019kez}. See also more recent constructions based on Horndeski theories~\cite{Fernandes:2025fnz, Fernandes:2025eoc}. } Naturally, these results make contact with the longstanding program of research on regular black holes. Since the 1960s, various authors have \textit{postulated} black hole metrics with singularity-free (regular) cores and studied their properties at a kinematical level~\cite{Sakharov:1966aja,1968qtr..conf...87B,Poisson:1988wc,Dymnikova:1992ux,Hayward:2005gi, Frolov:2014jva}. While a surprising amount has been learned from this endeavour~\cite{Carballo-Rubio:2018pmi, Carballo-Rubio:2018jzw, Carballo-Rubio:2019nel, Carballo-Rubio:2019fnb,Franzin:2022wai,Ghosh:2022gka, Vagnozzi:2022moj,Pedrotti:2024znu,Calza:2024xdh,Calza:2024fzo, Davies:2024ysj, Borissova:2025msp, Coviello:2025pla,Carballo-Rubio:2025fnc}, the program has faced serious limitations due to the absence of robust physical theories that actually predict the existence of regular black holes. This is not for lack of trying, as many authors have attempted to embed regular black holes as solutions within a panoply of frameworks~\cite{Frolov:1989pf,Barrabes:1995nk,Easson:2002tg, Nicolini:2005vd,Olmo:2012nx,Balakin:2015gpq,Bazeia:2015uia,Chamseddine:2016ktu,Bambi:2016xme, Easson:2017pfe, Bejarano:2017fgz,Colleaux:2017ibe,Cano:2018aod, Colleaux:2019ckh,Guerrero:2020uhn,Brandenberger:2021jqs,Olmo:2022cui,Biasi:2022ktq,Junior:2024xmm,Ovalle:2024wtv,Alencar:2024yvh,Bronnikov:2024izh,Bolokhov:2024sdy,Skvortsova:2024wly,Ayon-Beato:1998hmi,Bronnikov:2000vy,Ayon-Beato:2000mjt,Bronnikov:2000yz,Ayon-Beato:2004ywd,Dymnikova:2004zc,Berej:2006cc,Balart:2014jia,Fan:2016rih, Bronnikov:2017sgg,Shojai:2022pdq,Junior:2023ixh,Murk:2024nod,Li:2024rbw,Zhang:2024ljd,Erices:2024iah,Estrada:2024uuu,Huang:2025uhv,Calza:2025mrt}. However, all of these approaches face serious drawbacks, such as requiring \textit{ad hoc} exotic matter, fine tuning of parameters, or the persistence of singularities in the general solution space. In the formalism of~\cite{Bueno:2024dgm}, regular black holes emerge as the unique exact solutions to the equations of motion, without fine tuning or \textit{ad hoc} matter. The result is perhaps the first comprehensive approach to regular black holes~\cite{Konoplya:2024hfg, DiFilippo:2024mwm, Konoplya:2024kih, Ma:2024olw, Frolov:2024hhe, Wang:2024zlq, Fernandes:2025fnz, Fernandes:2025eoc}. 

A more pressing question is that of realistic gravitational collapse in these models. Under which circumstances (if any) does the collapse of matter result in the singularity-free solutions obtained in~\cite{Bueno:2024dgm}? Initial progress on this was made in~\cite{Bueno:2024eig, Bueno:2024zsx}, where four of us (BCHM) addressed the problem of thin shell collapse in theories incorporating infinite towers of higher curvature corrections. There it was demonstrated that collapsing shells, upon reaching a minimum radius, will always undergo a bounce, emerging into a new universe through a white hole explosion. The evolution is completely nonsingular and, from the exterior of the collapsing body, produces the regular black hole solutions of~\cite{Bueno:2024dgm}. 

While these results demonstrate that gravitational collapse \textit{can} occur in a nonsingular fashion, it is crucial to assess if this is robust to deformations of the matter model under consideration. In general relativity, gravitational collapse is an exceptionally rich phenomenon, even in the spherically symmetric sector --- see, {\it e.g.},~\cite{Joshi:2008zz} for an overview. Therefore, to better understand exactly which types of singularities the resummation of these classical corrections can resolve, it is necessary to explore more complex and realistic matter models. To this end, here we elaborate on nonsingular gravitational collapse by considering the Oppenheimer-Snyder analysis within the framework of~\cite{Bueno:2024dgm, Bueno:2024eig, Bueno:2024zsx}. Considering stars composed of pressureless dust, we will show that when infinite towers of higher-curvature terms are included in the action, the collapse is regular throughout.  Because the gravitational dynamics of the star interior is closely related to FLRW cosmology, we pause along the way to demonstrate that the cosmological solutions are generically nonsingular. If the series of corrections is truncated at finite order, then our work naturally ties in with existing studies of Oppenheimer-Snyder collapse with higher-curvature interactions, {\it e.g.}~\cite{Ilha:1996tc}. In these cases, the collapse remains singular, albeit less so than in general relativity. This highlights the fact that it is truly the \textit{full} resummation of higher-curvature terms that is required for singularity~resolution.

%
%

%


\section{Quasi-topological gravities}

Consider a general $D$-dimensional theory of gravity $\mathcal{L}(g^{ab}, R_{cdef})$ constructed from arbitrary contractions of the Riemann tensor and the metric, 
\begin{equation}
S=\frac{1}{16 \pi \GN} \int \mathrm{d}^D x \sqrt{|g|} \mathcal{L}(g^{ab}, R_{cdef})\,.
\label{eq:teogen}
\end{equation}
The general equations of motion of such a theory are given by \cite{Padmanabhan:2011ex}
\begin{equation}
P_{acde}R_{b}{}^{cde}-\frac{1}{2}g_{ab} \mathcal{L}+2 \nabla^c \nabla^d P_{acbd}=0\,, \quad P^{abcd}=\frac{\partial \mathcal{L}}{\partial R_{abcd}}\,.
\end{equation}
Observe that the term $\nabla^c \nabla^d P_{acbd}$ will generically introduce higher derivatives of the metric. Those theories for which $ \nabla^d P_{acbd}=0$ on general space-times possess second-order equations of motion and conform the class of Lovelock gravities \cite{Lovelock1,Lovelock2,Padmanabhan:2013xyr},
\be
\nabla^d P_{acbd}=0 \quad \Leftrightarrow \quad \text{Lovelock gravity}\,.
\ee
For a given space-time dimension $D$, Lovelock theories can be defined up to curvature orders $n \leq \lfloor D/2 \rfloor$ (integer part of $D/2$), being topological when $n=D/2$. As a consequence, for any finite $D$, it will not be possible to include in \eqref{eq:teogen} Lovelock theories of arbitrarily large curvature order. 

We will be interested in a generalization of Lovelock gravities which can be defined at any curvature order and space-time dimension $D \geq 5$. Consider a general spherically symmetric ansatz,
\begin{equation}\label{Nf}
\mathrm{d}s^2_{\rm SS}=-N(t,r)^2f(t,r)\mathrm{d}t^2+\frac{\mathrm{d}r^2}{f(t,r)}+r^2\mathrm{d}\Omega_{(D-2)}^2\, .
\end{equation} 
The theory $\mathcal{L}(g^{ab}, R_{cdef})$ will be a \emph{quasi-topological gravity} \cite{Oliva:2010eb,Quasi,Dehghani:2011vu,Ahmed:2017jod,Cisterna:2017umf,Bueno:2019ycr,Bueno:2022res,Moreno:2023rfl,Bueno:2024dgm} if
\be \label{SSs} 
\left.\nabla^d P_{acdb}\right|_{\rm SS}=0 \quad \Leftrightarrow \quad \text{Quasi-topological gravity}\,,
\ee
where $\vert_{ \rm SS}$ stands for evaluation on the SS ansatz \eqref{Nf}. We will denote a quasi-topological gravity formed by linear combinations of $n$-th order curvature invariants as $\mathcal{Z}_{(n)}$. In $D=4$, the unique quasi-topological gravity with non-trivial dynamics is Einstein gravity. For $D \geq 5$, instances of quasi-topological gravities up to quintic curvature order are provided by the following theories \cite{Bueno:2024eig,Bueno:2024zsx}:
\begin{subequations}\label{eq:Znexplicit}
\begin{align}
 \mathcal{Z}_{(1)}&=R\,, \\
 \mathcal{Z}_{(2)}&=\frac{\mathcal{Z}_{(1)}^2}{D(D-1)}-\frac{1}{(D-2)} \left [\frac{4 Z_{ab}Z^{ab}}{D-2}-\frac{W_{abcd} W^{abcd}}{D-3}\right]\,, \\
\nonumber \mathcal{Z}_{(3)}&=\frac{3\mathcal{Z}_{(1)}\mathcal{Z}_{(2)}}{D(D-1)}-\frac{2 \mathcal{Z}_{(1)}^3}{D^2(D-1)^2}+\frac{24}{(D-2)^2(D-3)} \left[\frac{ W\indices{_a_c^b^d} Z^a_b Z^c_d}{(D-2)}-
   \frac{   W_{a c d e}W^{bcde}Z^a_b}{(D-4)} \right]\\&+\frac{16
   Z^a_b Z^b_cZ_a^c}{(D-2)^4}   +\frac{2(2 D-3) W\indices{^a^b_c_d}W\indices{^c^d_e_f}W\indices{^e^f_a_b}}{ (D-2)(D-3) (D ((D-9)
   D+26)-22)} \,,   \\
\nonumber
\mathcal{Z}_{(4)}&=\frac{4\mathcal{Z}_{(1)}\mathcal{Z}_{(3)}-3 \mathcal{Z}_{(2)}^2}{D(D-1)}+\frac{96}{(D-2)^2(D-3)} \left[\frac{(D-1)\left ( W_{abcd} W^{abcd} \right)^2}{8D(D-2)^2(D-3)}-\frac{4Z_{a c} Z_{d e} W^{bdce} Z^{a}_b}{(D-2)^2(D-4)}\right. \\ \nonumber &-\frac{(2D-3) Z_e^f Z^e_f W_{abcd} W^{abcd}}{4(D-1)(D-2)^2}-
\frac{2 W_{acbd} W^{c efg} W^d{}_{efg} Z^{ab} }{D(D-3)(D-4)}  +\frac{(D^2-3D+3) \left (Z_a^b Z_b^a\right )^2}{D(D-1)(D-2)^3}\\& \left. -\frac{Z_a^b Z_b^c Z_c^d Z_d^a}{(D-2)^3}+\frac{(2D-1)W_{abcd} W^{aecf} Z^{bd} Z_{ef}}{D(D-2)(D-3)}\right]\,,\\ \nonumber
\mathcal{Z}_{(5)}&=\frac{5\mathcal{Z}_{(1)}\mathcal{Z}_{(4)}-2\mathcal{Z}_{(2)}\mathcal{Z}_{(3)}}{D(D-1)}+\frac{6 \mathcal{Z}_{(1)}\mathcal{Z}_{(2)}^2-8 \mathcal{Z}_{(1)}^2\mathcal{Z}_{(3)}}{D^2(D-1)^2}+\frac{768 Z_a^b Z_b^c Z_c^d Z_d^e Z_e^a}{5(D-2)^6(D-3)(D-4)} \\ \nonumber & + \frac{24(D-1)W_{ghij} W^{ghij}W\indices{_a_b^c^d}W\indices{_c_d^e^f}W\indices{_e_f^a^b} }{D(D-2)^3 (D-3)^2(D^3-9 D^2+26D-22)}-\frac{96(3D
-1)W^{ghij} W_{ghij}  W_{a c d e}W^{bcde} Z^a_b}{10D(D-1)(D-2)^4(D-3)^2(D-4)} \\ \nonumber & -\frac{768(D^2-2D+2)Z_a^b Z_b^a Z_c^d Z_d^e Z_e^c}{D(D-1)^2(D-2)^6(D-3)(D-4)}  -\frac{96(3D-1)(D^2+2D-4)W^{ghij} W_{ghij} Z_c^d Z_d^e Z_e^c}{D(D-1)(D+1)(D-2)^6(D-3)(D-4)}\\ \nonumber & +\frac{24(15D^5-148 D^4+527 D^3-800 D^2+472D-88)W\indices{_a_b^c^d}W\indices{_c_d^e^f}W\indices{_e_f^a^b} Z_{g}^h Z_h^g}{D(D-1)(D-2)^3(D-3)(D-4)(D^5-15D^4+91 D^3-277 D^2+418D-242)} \\\nonumber & +\frac{960 (D-1)}{(D-2)^4(D-3)^2} \left[\frac{(5D^2-7D+6)Z_g^h Z_h^g W_{abcd} Z^{ac} Z^{bd}}{10D(D-1)^2(D-2)}-\frac{Z_{a}^b Z_{b}^{c} Z_{cd} Z_{ef} W^{eafd}}{(D-1)(D-2)} \right. \\ \nonumber &- \frac{2(3D-1)Z^{ab} W_{acbd} Z^{ef}  W\indices{_e^c_f^g} Z^d_g}{D(D^2-1)(D-4)} +\frac{(D-3)W_{a c d e}W^{bcde} Z^a_b Z_f^g Z_g^f}{5D(D-1)^2(D-4)}\\  &\left. -\frac{(D-2)(D-3)(3D-2) Z^a_b Z^b_c W_{daef} W^{efgh} W_{gh}{}^{dc}}{4(D-1)^2(D-4)(D^2-6D+11)}+\frac{W_{ghij}W^{ghij} Z^{ac}Z^{bd}W_{abcd}}{20D(D-1)^2}\right]\,,
\end{align}
\end{subequations}
where $W_{abcd}$ stands for the Weyl curvature tensor and $Z_{ab}=R_{ab}-\frac{1}{D}R g_{ab}$ is the traceless Ricci tensor. Quasi-topological gravities of higher curvature order may be built with the aid of the following recursion formula \cite{Bueno:2019ycr}:
\begin{equation}
\label{eq:zrec}
\mathcal{Z}_{(n+5)}=\frac{3(n+3)\mathcal{Z}_{(1)}\mathcal{Z}_{(n+4)}}{D(D-1)(n+1)}-\frac{3(n+4)\mathcal{Z}_{(2)}\mathcal{Z}_{(n+3)}}{D(D-1)n}+\frac{(n+3)(n+4)\mathcal{Z}_{(3)}\mathcal{Z}_{(n+2)}}{D(D-1)n(n+1)}\, .
\end{equation}
As it turns out, quasi-topological gravities constructed from  \eqref{eq:Znexplicit} and \eqref{eq:zrec} satisfy a Birkhoff theorem, as it will be explicitly checked afterwards. 
Despite forming a particular subclass of theories in the infinite-dimensional moduli space of higher-curvature gravities, any effective theory of gravity may be mapped into a quasi-topological gravity through perturbative field redefinitions of the metric \cite{Bueno:2019ltp}, so they actually capture the most general theory of gravity when restricting to such an effective-field-theory regime.

If one is interested in spherically symmetric backgrounds, all quasi-topological gravities of a given curvature order $n$ contribute equally to the subsequent equations of motion (up to an overall coupling constant). This degeneracy implies that it is sufficient to pick a single quasi-topological representative $\mathcal{Z}_{(n)}$ at any $n$ (such as those in \eqref{eq:Znexplicit} and \eqref{eq:zrec}) to study the spherically symmetric sector of the most general quasi-topological gravity. Therefore, we will be considering the following theory of gravity:
\begin{equation}\label{QTaction}
S= \frac{1}{16\pi \GN}\int \mathrm{d}^Dx \sqrt{|g|}\left[R+\sum_{n=2}^{\infty}\alpha_{n}\mathcal{Z}_{n}\right]\, ,
\end{equation}
where the $\alpha_n$ stand for coupling constants with dimensions of $\left[ \text{length} \right]^{2n-2}$. Observe that we are actually allowing for the inclusion of an infinite tower of higher-curvature corrections: this is fundamental to achieve singularity resolution in the subsequent black hole solution \cite{Bueno:2024dgm}.

\subsection{Quasi-topological gravities as two-dimensional Horndeski theories}

As it turns out, there is a deeper reason that justifies why quasi-topological gravities have second-order equations for spherically symmetric backgrounds. Indeed, consider the following $D$-dimensional spherically symmetric metric:
\begin{equation}\label{sphericmetric}
\mathrm{d}s^2=\gamma_{\mu\nu}\mathrm{d} x^{\mu}\mathrm{d}x^{\nu}+\varphi(x)^2 \mathrm{d}\Omega^2_{(D-2)}\, ,
\end{equation}
where $\gamma_{\mu\nu}\mathrm{d} x^{\mu}\mathrm{d}x^{\nu}$ stands for a two-dimensional Lorentzian metric (whose indices are denoted with Greek letters and run from 0 to 1) and $\varphi(x)$ is a scalar field that depends on the coordinates of the two-dimensional Lorentzian space. Performing a dimensional reduction of \eqref{QTaction} on top of \eqref{sphericmetric}, one ends up with the following two-dimensional theory \cite{Bueno:2024zsx,Bueno:2024eig}: 
\begin{equation}\label{eq:2daction}
S_{\rm 2d}=\frac{(D-2)\Omega_{(D-2)}}{16\pi G_{\rm N}}\int \mathrm{d}^{2}x\sqrt{|\gamma|} \mathcal{L}_{\rm 2d}(\gamma_{\mu\nu},\varphi)\, , \quad \text{where} \quad \Omega_{(D-2)}\equiv \frac{2\pi^{(D-1)/2}}{\Gamma\left[\tfrac{D-1}{2}\right]}
\end{equation}
is the volume of the $(D-2)$-sphere and
\begin{align}
\label{eq:horn}
\mathcal{L}_{\rm 2d}=G_{2}(\varphi, X)-\Box\varphi G_{3}(\varphi, X)+G_{4}(\varphi, X)R^{\rm 2d}-2G_{4,X}(\varphi, X)\left[(\Box\varphi)^2-\nabla_{\mu}\nabla_{\nu}\varphi\nabla^{\mu}\nabla^{\nu}\varphi\right]\, ,
\end{align}
where $X=\gamma^{\mu \nu} \partial_\mu\, \varphi \partial_\nu \varphi$, $R^{\rm 2d}$ denotes the Ricci scalar of $\gamma_{\mu \nu}$, $\nabla_\mu$ stands for the two-dimensional covariant derivative, $\Box$ is the corresponding Laplacian and the functions $G_i(\varphi,X)$ are defined as follows:
\begin{align}
G_{2}(\varphi, X)&=\varphi^{D-2}\left[(D-1)h(\psi)-2\psi  h'(\psi)\right]\, ,\\\label{G3form}
G_{3}(\varphi, X)&=2\varphi^{D-3}h'(\psi)\, ,\\
\label{eq:G4}
G_{4}(\varphi, X)&=-\frac{1}{2}\varphi^{D-2}\psi^{(D-2)/2}\int \mathrm{d}\psi \psi^{-D/2}h'(\psi)\, ,
\end{align}
where\footnote{In the expression for $G_{4}(\varphi, X)$, one must take the primitive that ensures
\begin{equation}
G_4(1,1)=\frac{1}{D-2}\left (1+ \sum_{n=2}^\infty n \alpha_n \right)\,.
\end{equation}} $\psi=\dfrac{1-X}{\varphi^2}$ and $h'(\psi)=\mathrm{d}h/\mathrm{d}\psi$. Also, we have used the notation $G_{i,X} \equiv \partial_X G_i$. Observe that all the information about the putative $D$-dimensional quasi-topological gravity \eqref{QTaction} from which \eqref{eq:horn} is derived is encoded in the \emph{characteristic function} $h(\psi)$, 
 \be \label{eom_psi}
h(\psi)\equiv \psi + \sum_{n=2}^{\infty}\alpha_n \frac{D-2n}{D-2}  \psi^n\, .
\ee
Written this way, \eqref{eq:2daction} matches exactly the form of a two-dimensional Horndeski  scalar-tensor theory \cite{Horndeski:1974wa}, which by definition possesses second-order equations of motion. As a consequence, quasi-topological gravities have second-order equations of motion on top of spherically symmetric configurations because their spherical sector is equivalent to a two-dimensional Horndeski theory. 

For the sake of convenience, let us choose some coordinates for $\gamma_{\mu \nu}$ in which
\begin{equation}
\gamma_{\mu \nu} \mathrm{d}x^\mu \mathrm{d}x^\nu =-N(t,r)^2 f(t,r) \mathrm{d}t^2+ \frac{\mathrm{d}r^2}{f(t,r)}\,,
\end{equation}
and fix $\varphi=r$. In turn, this implies that $X=f(t,r)$ and $\psi=\dfrac{1-f}{r^2}$. Including a stress-energy tensor $T_{a b}$ on top of the $D$-dimensional theory \eqref{QTaction}, the subsequent equations of motion of \eqref{eq:horn} are given by
\begin{align}
\label{eom1}
 \partial_r \left[r^{D-1}h(\psi) \right]&=\frac{16\pi G_{\rm N}}{(D-2)N^2 f} r^{D-2}T_{tt} \,,\\
 \label{eom2}
\partial_t f  &=-\frac{16\pi G_{\rm N}}{(D-2)h'(\psi)} rf T_{tr} \,  \,,\\
\label{eom3}
\partial_r N &=\frac{8\pi G_{\rm N} }{(D-2)h'(\psi)} r N\left(T_{rr}+\frac{1}{N^2 f^2}T_{tt}\right) \,.
\end{align}
These equations can be seen to be equivalent to the full set of equations of motion for the $D$-dimensional theory \eqref{QTaction} on spherically symmetric backgrounds \eqref{Nf}.

\subsubsection{Junction conditions}

Consider the $D$-dimensional space-time to be divided into two manifolds $\mathcal{M}_+$ and $\mathcal{M}_-$ with a common interface $\Sigma$, corresponding to a $(D-1)$-domain wall. We will be interested in gluing appropriately these space-times $\mathcal{M}_+$ and $\mathcal{M}_-$ across $\Sigma$, which will be required to be endowed with a certain surface stress-energy tensor. Such matching conditions are called generalized Israel 
junction conditions (as they generalize the original ones obtained in \cite{Israel:1966rt} for GR) and we proceed to review them next. 


On the one hand, the first junction condition adopts the same form as in GR: the induced metric $h_{AB}^+$ on $\Sigma$ computed from $\mathcal{M}_+$ must coincide with the induced metric $h_{AB}^-$ calculated from the $\mathcal{M}_-$ side:
\begin{equation}
h_{AB}^+=h_{AB}^-\,,
\end{equation}
where we are using capital Latin indices to denote boundary indices. Assuming this first junction condition is satisfied, define $h_{AB}=h_{AB}^+=h_{AB}^-$. Let $n_a$ be the unit normal vector to $\Sigma$, normalized as $n_a n^a= \varepsilon=\pm 1$. Under spherically symmetry, we may write 
\begin{equation}
h_{AB} \mathrm{d}x^A \mathrm{d}x^B=h_{\tau \tau} \mathrm{d}\tau^2+ \varphi(\tau)^2 \mathrm{d}\Omega_{(D-2)}^2\,.
\end{equation} 
From the two-dimensional perspective, $\Sigma$ corresponds to a curve with induced metric $\mathrm{d}s^2_h=h_{\tau \tau} \mathrm{d}\tau^2$. Choosing $\tau$ to be the proper time of the curve, one can always set $h_{\tau \tau}=-\varepsilon$. However, it will be convenient to keep $h_{\tau \tau}$ general for the moment, in order to examine properly its variational problem. 

On the other hand, the second junction condition does depend on the theory under consideration and will generically differ from that of GR. This condition relies heavily on the analytical derivation of those boundary terms that need to be added to \eqref{QTaction} to produce a well-posed variational problem. While the direct finding of these terms from the $D$-dimensional action \eqref{QTaction} is quite problematic, it becomes a feasible task when restricting to spherical symmetry and using the equivalent two-dimensional theory. Indeed, the total two-dimensional action with the appropriate boundary terms that gives rise to a well-posed two-dimensional variational problem is \cite{Padilla:2012ze,Bueno:2024eig,Bueno:2024zsx}:
\begin{equation}
S_{\rm 2d}^{\rm total}=S_{\rm 2d}+\frac{(D-2) \Omega_{(D-2)}}{16 \pi \GN} \int_\Sigma \mathrm{d}\tau \sqrt{\vert h_{\tau \tau} \vert} \left[F_3+2G_4 K+4 \Box^h \varphi \, F_{4,Y} \right]\,,
\label{eq:2dtot}
\end{equation}
where $K=\nabla_\mu n^\mu$ is the extrinsic curvature of the curve $\Sigma$, $\Box^h$ is the Laplacian operator on $\Sigma$ and
\begin{equation}
F_l=\int_0^{\varphi_n} G_l(\varphi, Y+\varepsilon x^2) \, \mathrm{d}x\,, \quad Y=h^{\tau \tau} \dot{\varphi}^2\,, 
\end{equation}
where $\varphi_n=n^\mu \partial_\mu \varphi$, $l=3,4$, $\dot{\varphi}=\mathrm{d}\varphi/\mathrm{d}\tau$ and $h^{\tau \tau}=1/h_{\tau \tau}$. Using the two-dimensional action \eqref{eq:2dtot}, the second junction condition for the theory \eqref{QTaction} on spherically symmetric metrics adopts the following form:
\begin{align}
\Pi_{AB}^--\Pi_{AB}^+=8\pi \GN S_{AB}\,,&  \quad \Pi_{AB}=\Pi_{\tau \tau}+\frac{g_{ij}}{D-2} \Pi_{ij}\,, \\ \Pi_{\tau\tau}=\frac{(D-2)\varepsilon}{2\varphi^{D-2}} F_3\,, & \quad g^{ij} \Pi_{ij}= -\frac{\varepsilon}{\varphi^{D-3}\dot \varphi} \frac{\mathrm{d}}{\mathrm{d} \tau} \left ( \varphi^{D-2} \Pi_{\tau \tau} \right)\,,
\end{align}
where $i,j$ stand for those components along the angular directions of \eqref{Nf} and $ \Pi_{AB}^\pm $ denotes the evaluation of $\Pi_{AB}$ from the corresponding side $\mathcal{M}_\pm$. 




\subsection{Regular black holes as vacuum solutions}
 
Let us next examine how regular black holes arise as vacuum solutions of the theory of gravity \eqref{QTaction} featuring an infinite tower of higher-curvature corrections. Setting $T_{ab}=0$ in \eqref{eom1}, \eqref{eom2} and \eqref{eom3}, we find that
 \begin{equation}\label{vacuumeq}
 \partial_t f=0\, , \quad \partial_r N=0\, , \quad  \partial_r \left[r^{D-1}h(\psi) \right]=0\, .
 \end{equation}
The first equation above establishes that $f=f(r)$. The second equation imposes that $N=N(t)$. However, such a temporal dependence may always be removed via an appropriate time reparametrization, so we may safely set $N=1$. As a consequence, a Birkhoff theorem follows and the most general vacuum spherically symmetric solution of \req{QTaction} is also static and fully determined by a single function $f(r)$:
 \begin{equation}
\mathrm{d}s^2=- f(r) \mathrm{d}t^2+\frac{\mathrm{d}r^2}{f(r)}+r^2 \mathrm{d}\Omega^2_{(D-2)}\,,
\label{eq:ssans}
\end{equation}
whose equation of motion is given by the last equation in \req{vacuumeq}, which can be integrated into the following algebraic equation:
 \be \label{eom}
h(\psi)  = \frac{2\mathsf{M}}{r^{D-1}}  \, , 
\ee
where  $\mathsf{M}$ is an integration constant which is related to the ADM mass of the solution \cite{Arnowitt:1960es,Arnowitt:1961zz,Deser:2002jk}, $M$, through 
\begin{equation}\label{newM}
\mathsf{M} \equiv \frac{8\pi \GN M}{(D-2)\Omega_{(D-2)}}\, . 
\end{equation}
Solutions of \req{eom} correspond to deformations of the Schwarzschild-Tangherlini solution, which is recovered whenever $h(\psi)=\psi$ (that is, when the full tower of higher-curvature corrections is absent; \ie if $\alpha_n=0 \, \, \forall n$). If only a finite number of higher-curvature terms are introduced (\ie $\alpha_n=0$ for any $n>n_{\rm max}$, with $n_{\rm max} \in \mathbb{Z}^+$), the singularity is weakened with respect to the Einstein gravity case, but still always present, since:
\be 
f (r)= 1 - \left(\frac{2\mathsf{M} }{\alpha_{n_{\rm max} }} \right)^{1/n_{\rm max} } r^{2  - (D-1)/n_{\rm max} } + \cdots \, ,
\ee
near $r=0$. However,  when the infinite tower of corrections is included, the singularity is completely smoothed out \cite{Bueno:2024dgm} as long as the function $h(\psi)$ has an inverse for $\psi>0$ and the series that defines it (cf. \req{eom_psi}) has a finite radius of convergence. These conditions are met for quite general choices of the couplings $\alpha_n$, just imposing some mild conditions on them. By way of example, demanding $\alpha_{n}(D-2n)\ge 0\,\, \forall\, n$ and $\lim_{n\rightarrow\infty} |\alpha_{n}|^{\frac{1}{n}}=C>0$ are sufficient conditions to ensure regularity of the subsequent black hole solution. This fixes the following universal behavior for a regular black hole near $r=0$:
\begin{equation}
f(r)\overset{r \rightarrow 0}{ \simeq} 1-\frac{r^2}{C}+\dots
\label{eq:dscore}
\end{equation}
As a result, regular black holes develop a de Sitter core in its deep interior, whose radius is determined by the characteristic length scale of the infinite tower of higher-curvature terms. 

Next we will consider some illustrative examples (corresponding to specific choices for the couplings $\alpha_n$) in order to discuss explicit results. 



{\bf The Hayward solution}. For odd $D$, a particularly convenient choice for the couplings $\alpha_n$ is provided by
\begin{equation}\label{exI}
\alpha^{(1)}_n=\frac{D-2}{D-2n}\alpha^{n-1}\, ,\quad \text{which produces }  \quad h_{1}(\psi)=\frac{\psi}{1-\alpha\psi}\, ,
\end{equation}
where we take $\alpha$ to be a positive coupling with dimensions of [length]$^2$. Using \eqref{eom}, it turns out that the unique vacuum spherically symmetric solution of \eqref{QTaction} with the choice \eqref{exI} is given by
\begin{equation}\label{Haywardf}
f_{1}(r)=1-\frac{2\mathsf{M}r^2}{r^{D-1}+2\mathsf{M}\alpha}\, .
\end{equation}
Due to its formal resemblance with the four-dimensional Hayward black hole solution \cite{Hayward:2005gi}, we will refer to \req{Haywardf} as $D$-dimensional Hayward black hole (or simply as the Hayward black hole). 

Let us study the horizon structure of this solution. For the sake of concreteness, we will focus on the case $D=5$, which we will also concentrate on later in this manuscript. In this case, the zeros of $f_1$ are governed by the equation:
\begin{equation}
r^4-2\mathsf{M} r^2+2\mathsf{M} \alpha=0\,.
\end{equation}
The positive roots of this equation are:
\begin{equation}
r_{\pm}=\sqrt{\mathsf{M}\pm \sqrt{\mathsf{M}^2-2\mathsf{M}\alpha}}\,.
\end{equation}
We see that there can be at most two positive roots, so that the Hayward black hole in $D=5$ may have at most two horizons. This will be indeed the case whenever the mass $\mathsf{M}$ is above the critical value $\mathsf{M}_{\rm cr}=2\alpha$, so that $r=r_+$ corresponds to the location of the event horizon and $r=r_-$ represents an inner horizon. 

If $\mathsf{M}=\mathsf{M}_{\rm cr}$, then both horizons coincide and the resulting black hole is extremal, with a degenerate Killing horizon at $r_{\rm ext}=\sqrt{2 \alpha}$. Finally, if $\mathsf{M}<\mathsf{M}_{\rm cr}$, the corresponding solution is horizonless and represents a smooth gravitational soliton --- \ie a finite-energy, horizonless, geodesically complete configuration.

{\bf The modified Hayward solution}. The theory defined by \req{exI} is valid for odd space-time dimensions. In even dimensions, the choice \req{exI} is no longer possible as the density of order $n=D/2$ does not contribute to $h(\psi)$ (cf. \req{eom_psi}). A simple generalization of  \req{exI} valid for general even $D$ is given by \cite{Bueno:2024zsx}
\begin{equation}\label{hMH}
h_{\left(D/2\right)}(\psi)=\frac{\psi}{\left[1-(\alpha \psi)^{\frac{D}{2}}\right]^{\frac{2}{D}}}\, , 
\end{equation}
where $\alpha>0$ has units of length$^2$. From \req{eom}, one solves for the metric function $f(r)$ and finds
\begin{equation}\label{MHaywardf}
f_{(D/2)}(r)=1-\frac{2\mathsf{M}r^2}{  \left[r^{\frac{D(D-1)}{2}}+(2\mathsf{M}\alpha)^{\frac{D}{2}} \right]^{\frac{2}{D}}}\, .
\end{equation}
Let us analyze this solution in close detail for the particular case of $D=6$. In such a case, the solution reads
\begin{equation}
f_{(3)}(r)=1-\frac{2\mathsf{M} r^2}{[r^{15}+(2\mathsf{M}\alpha)^3]^{\frac{1}{3}}}\,.
\end{equation}
The solution will display horizons whenever there exist real positive roots of the algebraic equation:
\begin{equation}
r^{15}-8 \mathsf{M}^3 r^6+8 \mathsf{M}^3 \alpha^3=0\,.
\label{eq:reqmh}
\end{equation}
Defining $\rho=r^3$, one can reduce the previous equation into a quintic algebraic equation:
\begin{equation}
\label{eq:rhoeq}
\rho^5-8 \mathsf{M}^3 \rho^2+8 \mathsf{M}^3 \alpha^3=0\,.
\end{equation}
The discriminant of this equation may be computed to be $\Delta=-4096 \mathsf{M}^{12}\alpha^3 (6912 \mathsf{M}^{6}-3125 \alpha^9 )$. Define $\mathsf{M}_{\rm cr}=\frac{5^{5/6}}{2^{4/3}\sqrt{3}} \alpha^{3/2}$. We can distinguish the following cases:
\begin{itemize}
\item $\mathsf{M}>\mathsf{M}_{\rm cr}$. In this case, $\Delta<0$ and \eqref{eq:rhoeq} may be seen to admit two positive real solutions (the same follows for \eqref{eq:reqmh}). The solution will correspond to a regular black hole, with the largest positive root of \eqref{eq:reqmh} corresponding to the event horizon and the other positive root signaling the presence of an inner horizon.
\item $\mathsf{M}=\mathsf{M}_{\rm cr}$. In this case, $\Delta=0$ and \eqref{eq:rhoeq} admits a real and positive double root, with the remaining roots being either negative or complex (the same applies for \eqref{eq:reqmh}). The solution will correspond to a regular extremal black hole with a degenerate Killing horizon located at $r_{\rm ext}=\frac{5^{1/6}}{3^{1/6}} \sqrt{\alpha}$.
\item $\mathsf{M}<\mathsf{M}_{\rm cr}$. In this case, $\Delta>0$ and \eqref{eq:rhoeq} admits no positive roots. As a consequence, the solution will feature no horizon and will represent a gravitational soliton.

\end{itemize}

\subsection{FLRW solutions}

Let us now examine Friedmann-Lema\^{i}tre-Robertson-Walker (FLRW) solutions of the quasi-topological gravity \eqref{QTaction} composed of an infinite tower of higher-curvature terms. We consider the following cosmological ansatz with spherical sections:
\begin{equation}\label{ca}
\mathrm{d}s^2=-\mathrm{d}\tau^2+a(\tau)^2 \left[  \frac{\mathrm{d}\eta ^2}{1- \eta^2}+\eta^2\mathrm{d}\Omega_{(D-2)}^2\right]\, .
\end{equation}
As this ansatz fits precisely into the structure \eqref{sphericmetric}, it is clear that \eqref{QTaction} will have second-order equations of motion for the scale factor (following the nomenclature of \cite{Moreno:2023arp}, \eqref{QTaction} will also be a \emph{cosmological gravity}). As matter of fact, any quasi-topological gravity as defined in \eqref{SSs} will automatically be a cosmological gravity. 

We assume a perfect fluid permeates the spacetime, with stress-energy tensor given by
\begin{equation}
T_{ab}=(\rho+ p)u_a u_b + p g_{ab}\,,
\end{equation}
where $\rho$ is the density, $p$ the pressure and $u^a=\delta^a_\tau$ is the $D$-velocity of the fluid in the background \eqref{ca} ($\tau$ is assumed to be the proper time of the fluid). The subsequent Friedmann equation for the scale factor $a(\tau)$ takes the following compact expression:
\begin{equation}
h(\Phi)=\varrho\, , \quad \text{with} \quad \Phi  \equiv \frac{1+ \dot a(\tau)^2}{a(\tau)^2}\,,
\label{eq:friedprev}
\end{equation}
where $\dot{a}=\dfrac{\mathrm{d} a}{\mathrm{d}\tau}$ and
\begin{equation}
\varrho \equiv \frac{16\pi \GN \rho}{(D-2)(D-1)}\,.
\end{equation}
On the other hand, from the conservation equation for the stress-energy tensor of the fluid, one obtains
\begin{equation}
\dot \rho + (D-1) (\rho+ p ) \frac{\dot a(\tau)}{a(\tau)}=0\,.
\end{equation}

We can go no further without specifying an equation of state for the matter. Here, following the standard notation in cosmology, we shall explore a family of equations of states satisfying 
\be 
p = w \rho \, .
\ee
Here, $w$ is the equation of state parameter which is set by the particular type of matter under consideration. For example, for dust we will have $w = 0$. With the equation of state given, we can integrate the continuity equation to obtain,
\be 
\varrho = \frac{\mathsf{a}}{ a(\tau)^{(w+1)(D-1)}} \, ,
\label{eq:density}
\ee
for some constant $\mathsf{a}$. Therefore, the Friedmann equation \eqref{eq:friedprev} may be expressed as
\begin{equation}\label{Frie}
h(\Phi)=\frac{\mathsf{a}}{a(\tau)^{(w+1)(D-1)}}\,  , \quad \text{where} \quad \Phi  \equiv \frac{1+ \dot a(\tau)^2}{a(\tau)^2} \, .
\end{equation} 
Interestingly enough, for $w = 0$, this is formally equivalent to \eqref{eom} after the replacements $\Phi \leftrightarrow \psi$, $2\mathsf{M} \leftrightarrow \mathsf{a}$ and $r \leftrightarrow a(\tau)$. This will play an important role in the later sections. 

Before continuing to study the collapse of matter to form black holes, let us pause to elaborate in more detail on the cosmological consequences of infinite towers of quasi-topological gravities. It is natural to wonder whether the same conditions that result in the resolution of black hole singularities can also resolve other forms of singularities. This is also interesting for another reason. In any construction of regular black holes based on an effective stress tensor it is impossible to know how to appropriately generalize beyond the examples under consideration. For example, based on Hayward's original paper~\cite{Hayward:2005gi}, it is impossible to deduce what the ``Hayward Universe'' would look like. However, in the construction of~\cite{Bueno:2024dgm}, this is completely within reach. Since there is a unique theory that predicts, for example, the Hayward metric, it is possible study the implications of that theory for different spacetimes.  We will do this below, constructing along the way what is perhaps the first example of a Hayward cosmology. We begin with a short review of FLRW singularities in Einstein gravity. 

\paragraph{Einstein gravity cosmology.} In Einstein gravity, we have simply $h(\Phi) = \Phi$ and therefore the Friedmann equation reads,
\be 
\Phi = \frac{\mathsf{a}}{ a(\tau)^{(w+1)(D-1)}} \quad \Rightarrow \quad  \dot{a}^2 - \mathsf{a} a(\tau)^{2 - (w+1)(D-1)} = -1 \, .
\ee
This equation has the form of a particle moving in a one-dimensional potential with a negative total energy.  First, note that there exists a critical value of $w$ that determines the sign of the exponent in the potential,
\be 
w_\star \equiv - \left(\frac{D-3}{D-1} \right) \, .
\ee
If $w > w_\star$ then the exponent is negative, while if $w < w_\star$ it is positive. The latter situation corresponds to \textit{inflationary matter}, since the acceleration becomes positive,
\be 
\ddot{a}(\tau) = \frac{\left(w_\star - w \right) \mathsf{a}}{w_\star + 1} a(\tau)^{- \frac{2 w - w_\star + 1}{w_\star + 1}} \, .
\ee
For the sake of concreteness, let us focus on matter above the inflationary bound, \ie $w > w_\star$. In such a situation, the effective potential will always diverge to negative infinity for small values of $a(\tau)$. This directly reflects the existence of a big bang or big crunch singularity, with $a(\tau) \to 0$ and its derivatives blowing up.

\paragraph{Hayward cosmology.} Let us consider now how this situation is modified in resummed quasi-topological gravities. We focus first on the Hayward model, which is specified by the coupling constants given in eq.~\eqref{exI}. We then have the Friedmann equation for a Hayward universe,
\be 
\frac{\Phi}{1-\alpha \Phi} = \frac{\mathsf{a}}{ a(\tau)^{(w+1)(D-1)}} \quad \Rightarrow \quad \dot{a}(\tau)^2  -   \frac{\mathsf{a} \, a(\tau)^2}{\alpha \, \mathsf{a} + a(\tau)^{(w+1)(D-1)}} = -1 \, . 
\ee
Again, this equation has the form of a particle moving in a one-dimensional potential. It's easy to see that when the scale factor is large, the effective potential reduces to the Einstein gravity one. On the other hand, the behaviour for small scale factor is dramatically altered. There, so long as $w > -1$, we have
\be\label{eqn:friedmann_smalla} 
\dot{a}(\tau)^2 - \frac{a(\tau)^{2}}{\alpha} = -1 \, .
\ee
Thus, rather than diverging to negative infinity, the potential asymptotically goes to zero from below for small scale factor. As a consequence, the Hayward universe will generically undergo a bounce which replaces the big bang or big crunch singularity of Einstein gravity. 

We can see the bounce more explicitly by directly solving the Friedmann equation~\eqref{eqn:friedmann_smalla} in the vicinity of $a(\tau) = 0$. This is most easily accomplished by first differentiating \eqref{eqn:friedmann_smalla},
\be 
\ddot{a}(\tau) = \frac{a(\tau)}{\alpha} \, ,
\ee 
solving for $a(\tau)$ and then fixing the integration constants such that~\eqref{eqn:friedmann_smalla} holds.  The general solution reads,
\be\label{eqn:bouncy} 
a(\tau) = \sqrt{\alpha} \cosh \left(\frac{\tau - \tau_\mathrm{min}}{\sqrt{\alpha}} \right) \, .
\ee
This describes a bounce that occurs at $\tau  =  \tau_\mathrm{min}$, where the scale factor reaches a minimum value $a(\tau) = \sqrt{\alpha}$ before rebounding. Crucially, we can directly see that not only $a(\tau)$ but also all of its derivatives remain finite as $\tau \to \tau_\mathrm{min}$, reflecting the completely nonsingular nature of the Hayward universe. 

\paragraph{Cosmology for general resummations.} While we have focused above on the case of the Hayward cosmology, the qualitative features are completely general. Indeed, suppose we have chosen the coupling constants such that the characteristic polynomial is $h(x)$. Then, we can formally write the Friedmann equation for this theory as
\be\label{eqn:gen_friedmann} 
h(\Phi) = \frac{\mathsf{a}}{ a(\tau)^{(w+1)(D-1)}} \quad \Rightarrow \quad \dot{a}(\tau)^2 - a(\tau)^2 h^{-1} \left(\frac{\mathsf{a}}{ a(\tau)^{(w+1)(D-1)}} \right) = -1  \, ,
\ee
where $h^{-1}(x)$ is the \textit{inverse} of $h(x)$, \ie $h(h^{-1}(x)) = x$. 

Now, further suppose that the coupling constants have been chosen so that the corresponding theory admits regular black hole solutions. \textit{Sufficient} conditions for this were described in~\cite{Bueno:2024dgm} and, for the asymptotically flat case, read
\be 
(D-2n) \alpha_n \ge 0  \, \, \forall n \quad \text{and} \quad \lim_{n \to \infty} |\alpha_n|^{1/n} = C > 0 \, .
\ee
When these sufficient conditions are met, the function $h(x)$ will have radius of convergence $1/C$ and, because of the first condition, will diverge at $x = 1/C$. In turn, this implies that 
\be 
\lim_{x\to\infty} h^{-1}(x) = \frac{1}{C} \, .
\ee

Consider then the general Friedmann equation~\eqref{eqn:gen_friedmann}.  We find that provided $w > -1$ the argument of $h^{-1}(x)$ tends to infinity as $a(\tau) \to 0$. Hence, we have from the above considerations that the small $a(\tau)$ behaviour of the Friedmann equation is completely universal and model independent,
\be 
\dot{a}(\tau)^2 - \frac{a(\tau)^2}{C} = -1 \, .
\ee
This is identical to the Hayward model discussed above for $C = \alpha$, and hence the solution is again given by~\eqref{eqn:bouncy} and describes a bounce. Thus, the same sufficient conditions that ensure the regularity of the black hole geometries \textit{also} imply the existence of bouncing cosmologies.

\section{Oppenheimer-Snyder collapse}
In this section we consider the collapse of a spherical star of pressureless dust\footnote{As it turns out, the perfect fluid inside the star must necessarily be pressureless in this construction (see \emph{e.g.}, \cite{Smoller1994}). Indeed, the outside solution has vanishing stress-energy tensor, while the interior solution harbors a perfect fluid. As a consequence, no flow of energy may exist between from the interior to the exterior, which is achieved only if pressure is exactly zero (so that the gravitational equations are not discontinuous across the interface $\Sigma$).} in a general quasi-topological theory.  In the GR context, this is the well-known setup originally considered by  Oppenheimer and Snyder \cite{Oppenheimer:1939ue}. 

In the interior of the star, the metric is given by  \req{ca}, where the scale factor satisfies the Friedmann equation \req{Frie} with $w = 0$, corresponding to dust. On the other hand, the exterior solution is given, by virtue of Birkhoff's theorem, by \req{eq:ssans}, where $f(r)$ is determined by \req{eom}.  The induced metric on the star surface as seen from the outside is given by
\begin{equation}
\diff s^2_+=-\left(f(R(\tau)) \dot T(\tau)^2 - \frac{\dot R(\tau)^2}{f(R(\tau))}\right) \diff \tau^2 + R(\tau)^2 \diff \Omega_{(D-2)}^2\, ,
\end{equation}
where we parametrized the surface by $r=R(\tau)$, $t=T(\tau)$. From the interior perspective, the surface is parametrized by $\eta=\eta_0$, so the induced metric reads
\begin{equation}
\diff s^2_-=-\mathrm{d}\tau^2+a(\tau)^2 \eta_0^2\mathrm{d}\Omega_{(D-2)}^2\, ,
\end{equation}
where $\eta_0$ is the position of the star surface in the comoving frame. The first junction condition, $h_{AB}^+=h_{AB}^-$, forces both metrics to match. This yields
\begin{equation}\label{jun1}
f(R(\tau))^2 \dot T(\tau)^2=f(R(\tau))+\dot R(\tau)^2  \, , \quad  R(\tau)^2=a(\tau)^2 \eta_0^2\, .
\end{equation}
On the other hand, from the second junction condition we have
\begin{equation}
\Pi_{\tau\tau}^+ = \Pi_{\tau \tau}^- \, ,
\end{equation}
where we used the fact that $S_{AB}=0$, since no surface energy density on the interface is considered. For our theories, we have
\begin{equation}
\Pi_{\tau\tau}^{\pm}=\frac{(D-2)}{\varphi}\int_0^{n_{\pm}^\mu \partial_\mu \varphi_{\pm}} \diff z \, h'\left(\frac{1+\dot\varphi^2-z^2}{\varphi^2} \right)\, .
\end{equation}
where the normal vectors read, respectively,
\begin{align}
n_+=\frac{\dot R(\tau)}{f(R)} \partial_{t}+ f(R)\dot T(\tau) \partial_r\, ,\quad 
n_-=\frac{\sqrt{1-\eta_0^2}}{a(\tau)}\partial_\eta \, ,
\end{align}
and $\varphi_+=r$, $\varphi_-=a(\tau) \eta$. From this, we find
\begin{align}
n_{+}^\mu \partial_\mu \varphi_+=f(R(\tau))\dot T(\tau)\, ,\quad 
n_{-}^\mu \partial_\mu \varphi_-=\sqrt{1-\eta_0^2}\, .
\end{align}
Hence the second junction condition reads
\begin{equation}
\frac{(D-2)}{R}\int_0^{f \dot T} \diff z \, h'\left(\frac{1+\dot R^2 -z^2}{R^2} \right) = \frac{(D-2)}{R}\int_0^{\sqrt{1-\eta_0^2}} \diff z \, h'\left(\frac{1+\dot R^2 -z^2}{R^2} \right) \, ,
\end{equation}
where we already imposed the relation between $R(\tau)$ and $a(\tau)$ coming from the first junction condition. This simply reduces to
\begin{equation}
f(R(\tau))\dot T(\tau)= \sqrt{1-\eta_0^2}\, .
\end{equation}
We can use this expression to simplify the first equation in \req{jun1}. The result reads
\begin{equation}\label{geodesic}
\dot R(\tau)^2 +\eta_0^2= 1 -f(R(\tau))\,  .
\end{equation}
This is nothing but the equation for a timelike radial geodesic on the black hole background with energy $E^2=1-\eta_0^2$. Hence, for general quasi-topological theories, each point in the dust star  follows a geodesic, just like for Einstein gravity. Assuming $\dot R(0)=0$, \ie that the star has not started collapsing yet at $\tau=0$, this can be alternatively written as
\begin{equation}\label{geodesic1}
\dot R(\tau)^2 +f(R(\tau)) =f(R_0)\,  ,
\end{equation}
where we defined $R_0\equiv R(0)$, namely, the star initial coordinate radius.

Observe that using the relation between $R(\tau)$ and $a(\tau)$ \req{jun1} and the defining equation for $f(R)$ \req{eom} we can rewrite the above equation as
\begin{equation} \label{geo2}
h\left(\psi \right) =\frac{2\mathsf{M}}{\eta_0^{D-1}a(\tau)^{D-1}}\quad \text{where} \quad \psi=\frac{1-f(R(\tau))}{R(\tau)^2}=\frac{1+\dot{ a}(\tau)^2}{a(\tau)^2}=\Phi\, .
\end{equation} 
Hence, the Friedmann equation  \req{Frie} for the scale factor is equivalent to the geodesic equation followed by the dust particles at the surface of the star on the corresponding black hole background. Comparing both equations we find the relations
\begin{equation}
\mathsf{a}=\frac{2\mathsf{M}}{\eta_0^{D-1}}=  \frac{ \varrho R^{D-1}}{\eta_0^{D-1}} \, ,
\end{equation}
which are the same as for Einstein gravity \cite{Poisson:2009pwt}. Note that $\varrho$ is proportional to $R^{1-D}$, as it may be equivalently derived from \eqref{eq:density} after setting $w=0$. 
Naturally, $M$ can be interpreted in the present context as the mass of the black hole produced by the collapse of the star. This can be expressed in terms of the initial density, $\rho_0\equiv \rho(0)$, and radius of the star as
\begin{equation}\label{Mrho}
2\mathsf{M}=\varrho_0 R^{D-1}_0 \quad  \Leftrightarrow \quad (D-1)M= \Omega_{(D-2)} \rho_0 R^{D-1}_0\, .
\end{equation}
In sum, the master equation is \req{geodesic1}, where $f(R)$ can be fully expressed, for a given theory, in terms of the initial density of the star and its radius using \req{Mrho}.


\subsection{Einstein gravity}
In order to analyze in more detail the evolution of the star we may consider explicit models. The simplest case is of course Einstein gravity \cite{Oppenheimer:1939ue}. The radius of the dust star evolves according to the equation
\begin{equation}
\dot R(\tau)^2+\varrho_0R_0^2-\frac{\varrho_0 R_0^{D-1}}{R(\tau)^{D-3}}=0\, ,
\end{equation}
Starting at some $R_0> R_{\rm Sch}\equiv  \sqrt{\varrho_0}R_0^2$, $R(\tau)$ decreases monotonically and reaches zero size in a finite proper time. 
At that point, the density becomes divergent. In $D=5$, $R(\tau)$ and therefore also $\varrho(\tau)$ can be obtained explicitly and they take a particularly simple form, namely,
\begin{equation}
R(\tau)=R_0 \sqrt{1-\varrho_0 \tau^2 }\, , \quad \varrho(\tau)=\frac{\varrho_0 R_0^4}{R(\tau)^4}\, .
\end{equation}
At a proper time
\begin{equation}
\tau_{\rm Sch}=\sqrt{\frac{1}{\varrho_0}-R_0^2}
\end{equation}
the star reaches its Schwarzschild radius, 
and, from that moment on, it becomes a Schwarzschild black hole as seen from the outside.
The final point of the collapse takes place for 
\begin{equation}
\tau_{\star}=\frac{1 }{\sqrt{\varrho_0}}\, , \quad R(\tau_{\star})=0\, .
\end{equation}
This evolution is illustrated in Fig.\,\ref{fig:hay}. Nothing changes qualitatively for $D\neq 5$, although the explicit formulas become considerably more involved.

\subsection{Einstein-Gauss-Bonnet gravity}
As a warm up, let us consider first including a Gauss-Bonnet correction to the Einstein-Hilbert action, which entails truncating the series of quasi-topological corrections in \req{QTaction} at order $n_{\rm max}=2$. Similarly to the Einsteinian case, the collapse process will give rise to a singular black hole. This will always be the case as long as the number of corrections remains finite.

 In the case of Einstein-Gauss-Bonnet, the corresponding static and spherically  solution of mass $\mathsf{M}$ is given by the metric function \cite{Boulware:1985wk, Wheeler:1985nh}
\begin{equation}
f(r)=1-\frac{r^2}{2\alpha}\left(\sqrt{1+\frac{8\mathsf{M}\alpha}{r^{D-1}}}-1\right) \, ,
\end{equation}
which represents a horizonless solution only when $2\mathsf{M}\geq\alpha$ in five spacetime dimensions.

The equation for $R(\tau)$ reads
\begin{equation}
\dot R(\tau)^2 +  \frac{R_0^2}{2\alpha}\left(\sqrt{1+4\alpha\varrho_0}-1\right) =  \frac{R(\tau)^2}{2\alpha}\left(\sqrt{1+4\alpha\varrho_0\frac{{R_0}^{D-1}}{R(\tau)^{D-1}}}-1\right)\, .
\end{equation}

\begin{figure*}
\centering \hspace{0cm}
\includegraphics[width=0.492\textwidth]{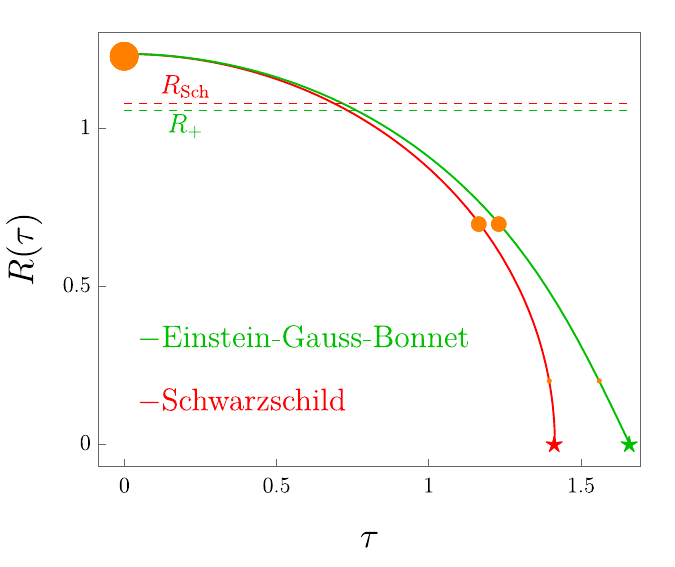}
\includegraphics[width=0.492\textwidth]{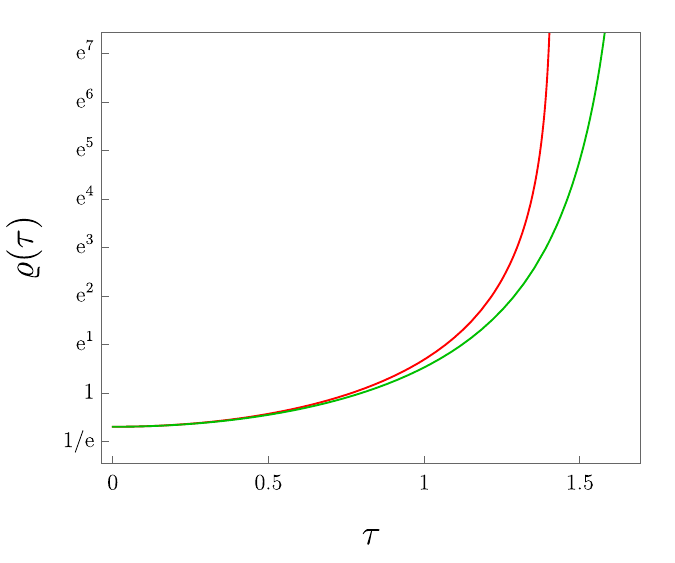}
\caption{We plot $R(\tau)$ for a collapsing star of orange dust in the case of $D=5$ Einstein gravity (red) and Einstein-Gauss-Bonnet gravity (green). We set $\varrho_0=1/2$, $\alpha=0.05$ and $R_0=1.237$.}
\label{fig:EGB}
\end{figure*}

For the black holes solutions, the dust star has a similar behaviour as in GR, reaching the zero size at a finite proper time
\begin{equation}
 \tau_\star=\int^{R_0}_{0}\frac{\sqrt{2\alpha}\,\diff x}{\sqrt{\sqrt{x^4+\frac{4\alpha\varrho_0 {R_0}^{D-1}}{x^{D-5}}}-x^2   - {R_0}^2 \left(\sqrt{1+4\alpha\varrho_0}-1\right)}}\, ,
\end{equation}
which can be numerically integrated. In the Fig\,.\ref{fig:EGB}, we present this process for specific parameters in five dimensions.

\subsection{Quasi-topological gravity}
Let us now see how this evolution gets modified in the case of quasi-topological gravities with regular black holes. First, we will discuss some generic physical aspects of Oppenheimer-Snyder collapse in theories \eqref{QTaction} including an infinite tower of higher-curvature corrections. Afterwards, we will consider explicit examples of collapse in $D=5$ and $D=6$ for specific choices of the couplings, respectively, but the conclusions are qualitatively the same for more general models and in higher-dimensions.

\subsubsection{General considerations}

Let us take a closer look at \eqref{geodesic1}. We note that it can be trivially reformulated as follows:
\begin{equation}
T+V=f(R_0)-1\,, \quad {\rm where} \quad T=\dot R(\tau)^2\,, \quad V=f(R(\tau))-1\,,
\end{equation}
which corresponds precisely to the trajectory of a test particle with total (negative) energy $f(R_0)-1$ moving in the potential $V$. Hence the effect of higher-curvature corrections is encoded in the modified gravitational potential $V$, which will no longer correspond to the Newtonian potential $V_{\rm N}=-\dfrac{2 \mathsf{M}}{R}$ appearing in Einstein gravity as soon as higher-curvature terms are introduced. 

\begin{figure*}
\centering \hspace{0cm}
\includegraphics[width=0.55\textwidth]{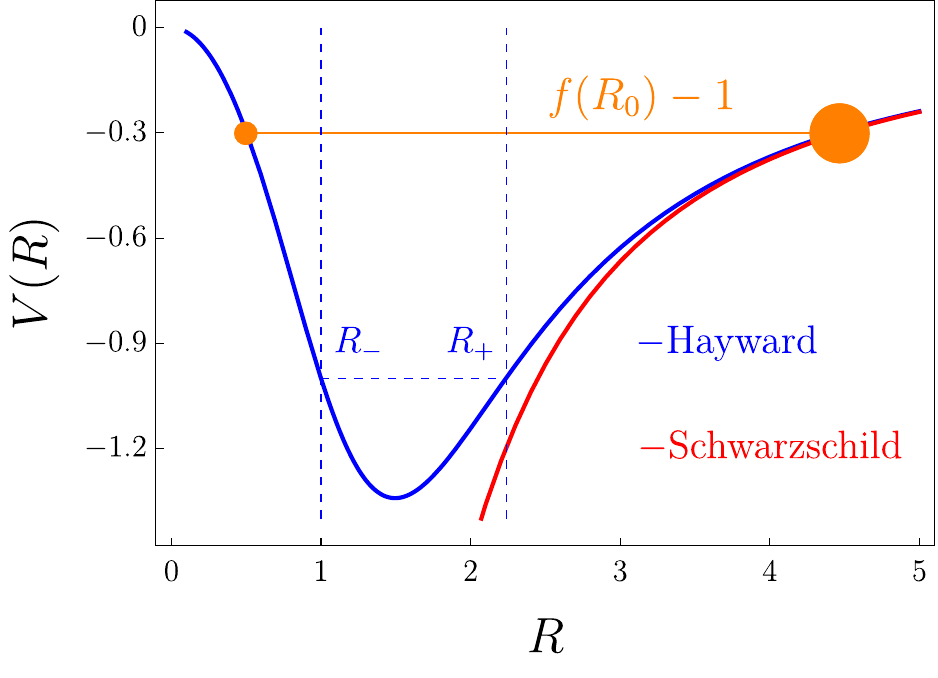}
\caption{Potential $V(R)=f(R)-1$ for the theory \eqref{exI} with $\alpha=5/6$. We have chosen $R_0 = 4.44$ and $\varrho_0 = 0.015$, resulting in a (non-extremal) Hayward black hole \eqref{Haywardf} in $D=5$ with $\mathsf{M}=3$ (all quantities expressed in some given units). Although we are plotting a particular case, the generic qualitative features replicate for all those quasi-topological gravities with regular black holes. The orange line stands for the energy $f(R_0)-1$ of the particle. It starts from rest at the right intersection of the orange line with $V(R)$, gets into the black hole interior once $r<r_+$ and, after surpassing the inner horizon at $r=r_-$, it reaches a turning point from which the particle is bounced back towards its initial position from which the motion started.
}
\label{fig:pothay}
\end{figure*}

From this observation, we also obtain a useful qualitative criterion to assess whether Oppenheimer-Snyder collapse in theories \eqref{QTaction} may give rise to regular black holes. In fact, since the potential is essentially fixed by the metric function $f(R)$ determining the black hole solution, whenever $f(R)$ corresponds to a regular configuration --- so that $f(R)$ is bounded, has the proper asymptotics and behaves near $r=0$ as $f(R \rightarrow 0)\simeq 1-\frac{R^2}{C}$, for $C>0$ (see \eqref{eq:dscore}) ---, we conclude that a particle starting with some negative energy $f(R_0)-1$ will have a turning point and will bounce. Consequently, we deduce that Oppenheimer-Snyder collapse will indeed result in the regular black holes specified by the solution $f(R)$ (see Fig. \ref{fig:pothay}). This expectation will eventually be confirmed with the examples to be shown afterwards.


We can also provide analytical arguments proving that a bounce will generically occur whenever the black hole solution that is being used is regular. First, differentiate \eqref{geodesic1} with respect to $\tau$ to arrive at:
\begin{equation}
\ddot{R}(\tau)=-\frac{f'(R(\tau))}{2}\,.
\label{simpleddotR}
\end{equation}
This extremely simple equation is valid for any choice of couplings in \eqref{QTaction}. Now, near $R=0$, the metric function $f(R)$ has the  universal behavior $f(R)\simeq 1-\frac{R^2}{C}$ for $C>0$, so we can approximate \eqref{simpleddotR} as:
\begin{equation}
\ddot{R}\simeq \frac{R}{C} \quad\Rightarrow\quad R=\sqrt{C}\sqrt{1- f(R_0)} \cosh\left ( \frac{\tau-\tau_{\rm min}}{\sqrt{C}}\right)\,, 
\label{aproxcosh}
\end{equation}
where $\tau_{\rm min}$ denotes the proper time at which the minimum is reached, $R(\tau_{\rm min})\equiv R_{\rm min}$,  and the overall factor $\sqrt{C}\sqrt{1- f(R_0)}$ is fixed upon demanding that \eqref{geodesic1} holds.  Consequently, if the particle is sufficiently near $R=0$, the approximation \eqref{aproxcosh} will  work effectively and the particle will follow a close-to-parabolic behaviour, going through a minimum $R_{\rm min}$ for $\tau=\tau_{\rm min}$. Then, $R$ will increase in a symmetrical way (with respect to the minimum) up to the point at which  the approximation \eqref{aproxcosh} is no longer valid.  Then the particle will rebound to its initial position, from where it was launched at rest. The correctness of this picture will be explicitly verified in the examples below.


\begin{figure*}
\centering \hspace{0cm}
\includegraphics[width=0.492\textwidth]{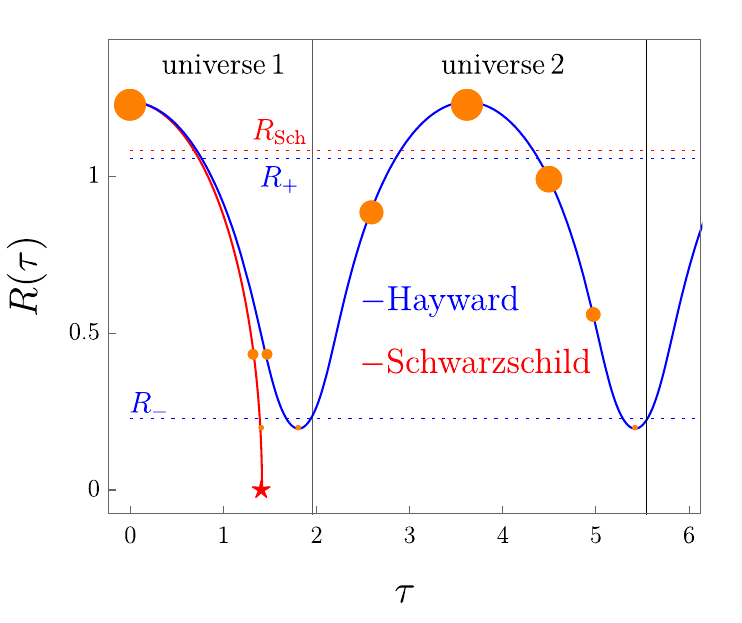}
\includegraphics[width=0.492\textwidth]{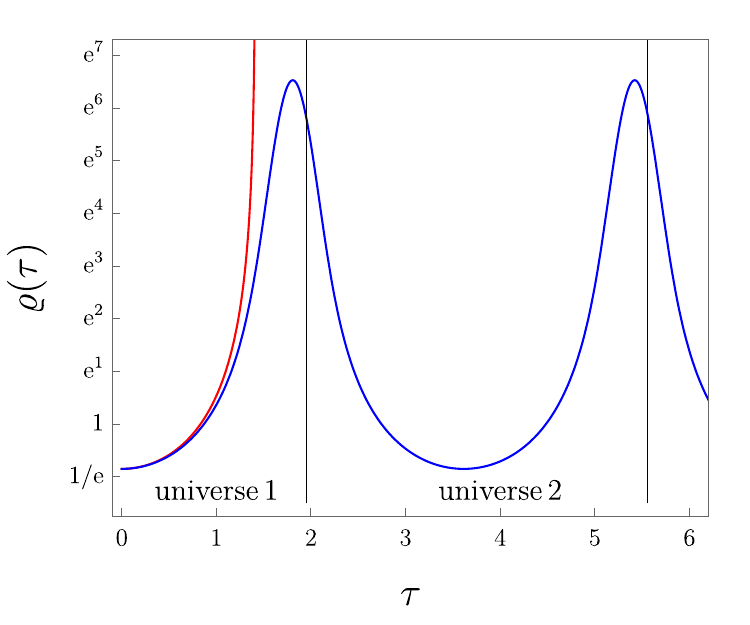}
\caption{(Left) We plot $R(\tau)$ for a collapsing star of orange dust in the case of $D=5$ Einstein gravity (red) and a quasi-topological gravity theory of the form \req{QTaction} with $\alpha_n=\tfrac{3}{5-2n}\alpha^{n-1}$ (blue). We set $\varrho_0=1/2$, $R_0=1.237$ and $\alpha=0.05$.
(Right) We plot the star density for the same models. In the case of Einstein gravity, the star collapses reaching zero size and infinite density after a finite proper time (red star). It leaves behind a Schwarzschild black hole, as seen from the exterior (the red dashed line corresponds to its Schwarzschild radius). On the other hand, for the quasi-topological theory, the star collapses forming outer and inner horizons (blue dashed lines in the left plot) corresponding to a Hayward black hole. Inside the inner horizon it  reaches a finite minimum size (and a maximum finite density), bouncing back and  emerging through a white hole in a new universe. Eventually it reaches a maximum size and the process starts over.
}
\label{fig:hay}
\end{figure*}

\subsubsection{Hayward exterior in $D=5$}
 A particularly simple modification of the Einstein gravity result within our framework corresponds to the choice \req{exI}, for which the exterior solution is a $D$-dimensional Hayward black hole. In that case, the equation for $R(\tau)$ reads
\begin{equation}
\dot R(\tau)^2-\frac{\varrho_0 R_0^{D-1}R(\tau)^2}{\left[R(\tau)^{D-1}+\alpha\varrho_0 R_0^{D-1} \right]}+\frac{\varrho_0 R_0^{2}}{1+\alpha \varrho_0}=0\, .
\end{equation}
This can be easily integrated numerically. The result for a particular set of values of $\varrho_0,R_0$ and $\alpha$ is shown in the $D=5$ case in Fig.\,\ref{fig:hay}. We choose $\mathsf{M} > \mathsf{M}_{\rm cr}$, namely, $\varrho_0 R_0^4>4\alpha$, so that the outcome of collapse is a Hayward black hole with two horizons and $f(R_0)>0$, so that the initial size of the star is greater than the outer horizon radius.
After a finite proper time the star reaches a minimum size
\begin{equation}
R_{\rm min}=R_0\sqrt{\alpha \varrho_0}\, ,
\end{equation}
which is always smaller than both the outer and inner horizons of the black hole it leaves behind,
\begin{equation}
R_{\rm min} < R_- < R_+<R_0\, ,
\end{equation}
where
\begin{equation}
R_{\pm}=\sqrt{\frac{\varrho_0 R_0^4}{2}\pm\sqrt{\frac{\varrho_0 R_0^4}{2}\left(\frac{\varrho_0 R_0^4}{2}-2\alpha\right)}}\, .
\end{equation}
When $R=R_{\rm min}$, the density reaches a maximum
\begin{equation}
\varrho_{\rm max}= \frac{\varrho_0 R_0^4}{R_{\rm min}^4} \quad \Longleftrightarrow \quad \alpha\varrho_{\rm max} = \frac{1}{\alpha\varrho_{\rm min}}\, ,
\end{equation}
 and both magnitudes bounce back, as shown in Fig.\,\ref{fig:hay}. The radius grows until $R=R_-$ and at that point the star emerges through a white hole in a new universe. It keeps on growing, eventually reaching its original size again,
\begin{equation}
R_{\rm max}=R_0\, .
\end{equation}
At that point, $\dot R=0$ again and the process is restarted. The evolution of $R(\tau)$ is therefore periodic with a proper-time period given by
\begin{equation}
T=2\int_{R_{\rm min}}^{R_{\rm max} }\frac{\diff x}{\sqrt{\frac{\varrho_0 R_0^4 x^2}{x^4+\alpha\varrho_0 R_0^4}-\frac{\varrho_0 R_0^{2}}{1+\alpha \varrho_0}}}\, ,
\end{equation}
and it consists of a never ending series of contractions and expansions of the star. Each pair of these takes place in a new universe, in each of which the star leaves behind a regular black hole of the Hayward type.

\begin{figure*}
\centering \hspace{0cm}
\includegraphics[width=0.494\textwidth]{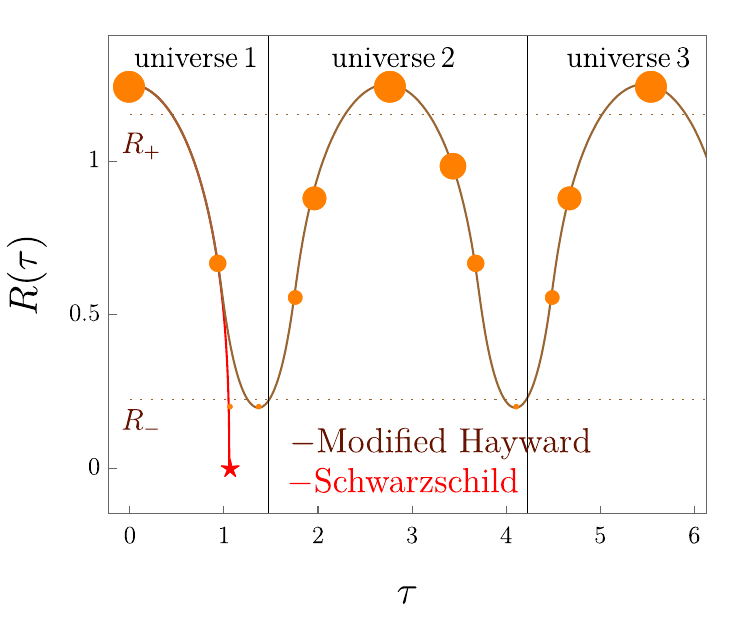}
\includegraphics[width=0.491\textwidth]{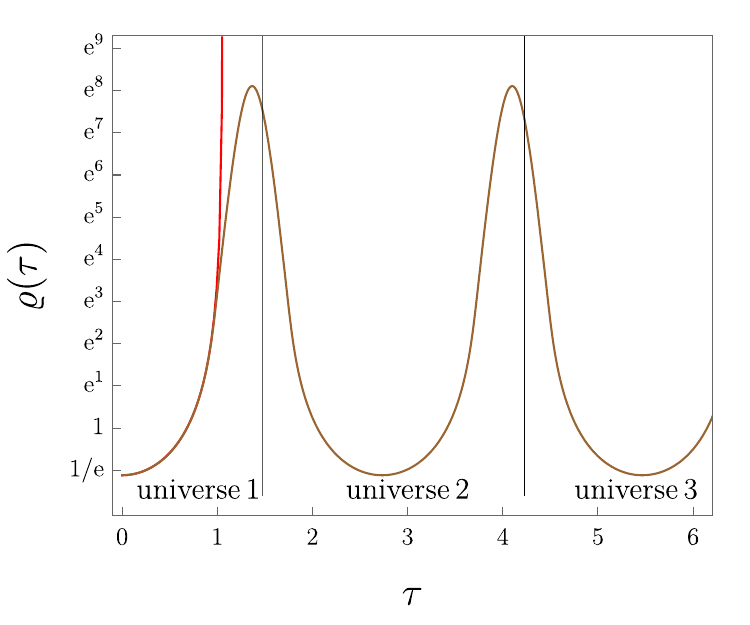}
\caption{(Left) We plot $R(\tau)$ for a collapsing star of orange dust in the case of $D=6$ Einstein gravity (red) and a quasi-topological gravity theory of the form \req{QTaction} with $\alpha_n$ defined by \req{hMH} and \req{eom_psi} (brown). We set $\alpha=0.05$, $R_0=1.25$, and $\varrho_0=1/2$.
(Right) Star density as a function of the proper time for the same models. 
}
\label{fig:mhay}
\end{figure*}

\subsubsection{Modified Hayward exterior in $D=6$}
As our second explicit example, let us consider the class of theories defined by \req{hMH} which, as opposed to the Hayward case, also exist in even dimensions. The simplest case corresponds to $D=6$, for which the exterior metric function reads
\begin{equation}
f_{(3)}(r)=1-\frac{2\mathsf{M}r^2}{  \left[r^{15}+(2\mathsf{M}\alpha)^{3} \right]^{\frac{1}{3}}}\, .
\end{equation}
In this case, the equation for the star radius reads
\begin{equation}
\dot R(\tau)^2-\frac{\varrho_0 R_0^5 R(\tau)^2}{ \left[R(\tau)^{15}+(\alpha\varrho_0 R_0^5)^{3} \right]^{\frac{1}{3}}}+\frac{\varrho_0 R_0^2}{[1+( \alpha \varrho_0)^3]^{\frac{1}{3}}}=0\, ,
\end{equation}
and the result of the numerical integration of this equation is displayed in Fig.\,\ref{fig:mhay} again in the case in which $\mathsf{M} > \mathsf{M}_{\rm cr}$, namely, $\varrho_0 R_0^5> \frac{5^{5/6}}{2^{1/3}\sqrt{3}}\alpha^{3/2}$, so that the outcome of the collapse process is a black hole with two horizons. Both $R(\tau)$ and $\varrho(\tau)$ behave in a qualitatively identical fashion to the five-dimensional case considered in the previous subsection.



\section{Conclusions}

That gravitational collapse might occur in a nonsingular fashion has been a topic of speculation for more than fifty years. However, a dearth of theoretical frameworks in which this idea can be explored concretely has hampered progress on assessing its feasibility. This has recently changed with the model introduced in~\cite{Bueno:2024dgm}, where it has been shown that incorporation of an infinite tower of higher-curvature corrections into the action generically leads to the resolution of the Schwarzschild singularity in any $D \ge 5$. Preliminary indications, based on the collapse of thin shells in these models, suggest that not only are the eternal black holes nonsingular in these theories, but gravitational collapse is as well~\cite{Bueno:2024eig, Bueno:2024zsx}. 

Our work has extended this program in two directions. First, we have illustrated that exactly the same theories that regularize black hole singularities also eliminate the FLRW cosmological singularities of general relativity, replacing them with cosmological bounces. Hence, the mechanism of singularity resolution via an infinite tower of corrections is thereby made more robust. Second, our main result concerned the gravitational collapse of dust \textit{{\`a} la} Oppenheimer and Snyder. In this context, we have shown that dust collapse is nonsingular---instead of terminating at a singularity, the collapsing star reaches a point of maximum (finite) density, after which a bounce occurs. This occurs generically, in the sense that any resummation of quasi-topological higher-curvature terms for which the corresponding static black hole solution is regular will yield a nonsingular Oppenheimer-Snyder collapse. Taken together with the results for thin shells, this strongly hints that the mechanism has a broad scope. 

In the same vein, the most important future directions concern determining exactly how generally collapse is nonsingular in these models.\footnote{A more comprehensive list of future directions appears in the discussion of~\cite{Bueno:2024zsx}.} The main limitation of this framework is that it is restricted to spherical symmetry but, happily, the problem of gravitational collapse is already rather nontrivial in this realm. Perhaps the simplest extension of our results here would be to consider \textit{inhomogeneous} dust. In general relativity, the relevant metrics belong to the Lema{\^i}tre-Tolman-Bondi class~\cite{Lemaitre:1931zz,Tolman:1934za, Bondi:1947fta} and played a key role in developing the mathematical formulation of weak cosmic censorship. In particular, the Lema{\^i}tre-Tolman-Bondi solutions can develop naked singularities~\cite{Christodoulou:1984mz}, and it would be interesting to assess whether the infinite tower of corrections eliminates these pathologies. A more complicated problem would be to consider collapsing matter with pressure~\cite{Joshi:2008zz}. Already in general relativity these configurations  require perturbative and/or numerical techniques, and so it is highly likely similar methods will be needed. One starting point for this problem would be to extend the Buchdahl theorem~\cite{Buchdahl:1959zz} to the theories considered here, which would allow one to assess when internal pressures become  insufficient to prevent gravitational collapse~\cite{buchdahl_appear}. 

More generally, one could consider explicit models for the collapsing matter, such as scalar or electromagnetic fields. However, one must take care in interpreting the results of such studies since it is somewhat unnatural to consider, for example, minimally coupled two-derivative matter when the gravitational sector is effectively a resummed derivative expansion. Nonetheless, on pragmatic grounds this opens the door to a number of further interesting problems~\cite{hennigar_appear}. For the matter of choice, one could study the collapse problem and also analyse the evolution of higher-derivative black hole entropy~\cite{Davies:2023qaa, Hollands:2024vbe, Wall:2024lbd} during the formation and evolution of black holes. A particularly interesting case would be the analysis of critical collapse~\cite{Choptuik:1992jv}. In general relativity, critical collapse provides a robust mechanism by which one may generate counter-examples to weak cosmic censorship. However, close to the critical threshold, the dynamics will depend sensitively on higher-order corrections, calling into question the robustness of these conclusions for a `more fundamental' theory. Despite this, there has been rather little investigation of critical collapse in the presence of higher-derivative interactions~\cite{Golod:2012yt, Deppe:2012wk}. It is therefore interesting to assess the impact of these terms on the process and, in particular, assess whether naked singularities are prevented. 

While all the directions discussed above implicitly assume working in five or more dimensions, it may also be possible to extend some lines  to the astrophysically relevant case of four dimensions. For example, in the cosmological setting there exist four-dimensional theories with second-order equations of motion at all orders in curvature, the resummation of which eliminates FLRW singularities~\cite{Arciniega:2018tnn, Moreno:2023arp}. Extending the results and prospective lines of research about collapse to four dimensions is less clearcut in pure gravity. However, including a scalar field and studying infinite towers of Horndeski lagrangians may lead to a framework wherein collapse problems can be tackled~\cite{Fernandes:2025fnz}. 


A deeper question going beyond all of this is understanding exactly what, at a fundamental level, is responsible for singularity resolution in these models and how far it can be pushed. Which features (if any) of top-down constructions are being captured by the infinite resummation of higher-order terms? Drawing an analogy with string theory, could these corrections be accounting for the finite size/non-local effects of strings as would result from resumming the $\alpha'$ corrections? We do not have answers to these questions, but hope to contribute to their understanding in the future.

%
%
%

\subsection*{Acknowledgements}
PB was supported by a Ramón y Cajal fellowship (RYC2020-028756-I), by a Proyecto de Consolidación Investigadora (CNS 2023-143822) from Spain’s Ministry of Science, Innovation and Universities, and by the grant PID2022-136224NB-C22, funded by MCIN/AEI/ 10.13039/501100011033/FEDER, UE.
The work of PAC received the support of a fellowship from “la Caixa” Foundation (ID 100010434) with code LCF/BQ/PI23/11970032.  \'AJM  was supported by a Juan de la Cierva contract (JDC2023-050770-I) from Spain’s Ministry of Science, Innovation and Universities. AVC was supported by a scholarship of the  CEX-2019-000918 project for ``Unidades de excelencia Mar\'ia de Maeztu''  funded by the Spanish Research Agency, AEI/10.13039/501100011033.


\bibliographystyle{JHEP-2}
\bibliography{Gravities.bib}

\end{document}